\documentclass{cimento1}
\usepackage{graphicx,color,epsfig,epsf,rotating,caption}
\title{Monte Carlo generators for top quark physics at the LHC}
\author{B.S.~Acharya\from{i11},
	F.~Cavallari\from{i2}, 
	G.~Corcella\from{i5}\from{i51}\thanks{\small 
Talk given by G.~Corcella at `V Workshop Italiano sulla Fisica $pp$ a LHC',
Perugia, Italy, 30 January -- 2 February, 2008.},
	R.~Di Sipio\from{i4} \atque
	G.~Petrucciani\from{i51}
	}

\instlist{ 
	\inst{i11} Abdul Salam International Center for Theoretical Physics 
and INFN, Sezione di Trieste, Italy
 	\inst{i2} INFN, Sezione di Roma, Italy
	\inst{i5} Museo Storico della Fisica e 
Centro Studi e Ricerche E.~Fermi, Roma, Italy
	\inst{i51} Scuola Normale Superiore and INFN, Sezione di Pisa, Italy
 	\inst{i4} Universit\`a di Bologna and INFN, Sezione di Bologna, Italy
}

\PACSes{
\PACSit{14.65.Ha }{Top quarks}\PACSit{12.38.Bx }{Perturbative calculations}
}

\begin{document}
\maketitle
\begin{abstract}
We review the main features of Monte Carlo generators for
top quark phenomenology and present some results for
$t\bar t$ and single-top signals and backgrounds at the LHC. 
\end{abstract}

\section{Introduction}
In order to perform precise measurements of properties of top quarks
at the LHC, 
the use of reliable tools will be essential.
Extensive work has been carried out to improve parton-shower
and matrix-element Monte Carlo (MC) codes for $t\bar t$
and single-top signals and backgrounds.
Parton shower generators, such as HERWIG \cite{herwig} or
PYTHIA \cite{pythia}, 
simulate top production and decay using leading-order (LO)
matrix elements and describe multiple radiation 
in the soft or collinear approximation.
HERWIG showers satisfy angular ordering, valid in
the soft approximation after azimuthal averaging, while PYTHIA 
orders its cascades according to transverse momentum or virtuality,
with an option to veto non-angular-ordered emissions.
To simulate hard and large-angle radiation, both PYTHIA and
HERWIG have been provided with matrix-element corrections.
HERWIG splits the physical phase space into a soft/collinear region,
where one trusts parton showers, and a region, corresponding
to hard and large-angle radiation, where the exact amplitude is used.
Moreover, HERWIG corrects the `hardest-so-far' emission in the shower.
The PYTHIA implementation is somewhat different: the
parton shower approximation is used throughout the entire physical phase space
and the first emission is generated exactly.
HERWIG \cite{corsey1,cms} and PYTHIA \cite{norb} 
include matrix-element corrections
to top-quark decay, but not to top production. 
Such corrections are also applied to $W$-boson production, which 
is matched to the $W+1$ 
jet process \cite{corsey2,miu}, one of the backgrounds for top
production.
At the end of the parton cascade, HERWIG and PYTHIA 
simulate the hadronization transition according to the cluster \cite{cluster} and
string \cite{string} models, respectively. As for the underlying event,
both programs deal with it by allowing multiple interactions \cite{multiple,jimmy}.
For this purpose, HERWIG is interfaced with the JIMMY code.
HERWIG++ \cite{hwpp} and PYTHIA 8 \cite{py8} 
are the new object-oriented versions of the two codes, 
written in C++. PYTHIA 8 implements only  
transverse-momentum evolution, with initial/final-state radiation and
multiple interactions interweaved in a common ordering.
HERWIG++ includes a mass-dependent term in the splitting functions, 
leading to a better treatment of the radiation off heavy quarks and heavy-hadron
fragmentation.

Besides HERWIG and PYTHIA, other parton showers generators are 
available. The ARIADNE \cite{ariadne} code uses PYTHIA hard scattering and hadronization,
but its showers are dipole cascades, i.e. $2\to 3$ branchings in the soft limit.
The evolution variable of ARIADNE is transverse momentum: in 
fact, implementing dipole emissions, there is no need for azimuthal averaging
and angular ordering. 
The SHERPA \cite{sherpa} generator implements showers which, like in
PYTHIA, are ordered in virtuality, with an option to reject 
non-angular-ordered emissions. Hadronization occurs following the
string model.
A feature of ARIADNE and SHERPA is that matrix-element
matching is implemented according to the CKKW--L \cite{ckkw,ll} 
method.
Jets are clustered according to the $k_T$-algorithm \cite{kt}, 
with a threshold $y_{\mathrm{cut}}$ on the $k_T$-clustering variable $y_{ij}$.
Exact matrix elements, weighted by the
Sudakov form factors, and parton-shower approximations are then used
above and below $y_{\mathrm{cut}}$.

The programs mentioned so far, even after matrix-element matching, still
yield LO rates. Next-to-leading order (NLO) 
cross sections and observables are instead given by the
codes MC@NLO \cite{mcnlo} and POWHEG \cite{powheg}, both available for top production.
MC@NLO implements the hard scattering at NLO and employs HERWIG for
showers and hadronization: its drawbacks are its
shower-model (HERWIG) dependence and its simulation of events with
negative weights. Such problems have been overcome by POWHEG, which
can be interfaced to any shower model and does not have any 
negatively-weighted event. The first POWHEG emission is the hardest in
transverse momentum $(p_T)$ and is generated exactly at NLO; the
subsequent cascade is ordered in $p_T$. To include angular ordering, 
one would need truncated showers, whose implementation is in
progress.

Finally, in order to simulate multi-jet final states in top-quark events 
and backgrounds, such as $W/Z+$ jets, the best available tools are the
so-called `matrix-element generators', such as ALPGEN \cite{alpgen}, 
MadGraph/MadEvent \cite{madg},
CompHep \cite{comphep}, HELAC \cite{helac}, GR@PPA \cite{grappa}
and WHIZARD \cite{whiz},
which include LO multi-parton 
amplitudes. Such programs can be interfaced to HERWIG or PYTHIA for
showering and hadronization, according to the Les Houches accord.
The ALPGEN code, for example, includes the following processes
($V=W$ or $Z$): $t\bar t +6$~jets, $V+6$~jets, $Wb\bar b+6$~jets,
$VV+3$~jets. 
It uses the so-called MLM prescription to match its partons with
the subsequent showers. Roughly speaking, one starts, e.g., with $W+n$ partons and defines
a cone jet algorithm. After showering, a 
$k$-jet event is accepted only if $k\geq n$ and each of the $n$ original 
partons is clustered into a different jet.
\begin{figure}[ht!]
\centerline{\resizebox{0.355\textwidth}{!}
{\includegraphics{top_lhc_ptpair_cut.eps}}%
\hfill%
\resizebox{0.355\textwidth}{!}{\includegraphics{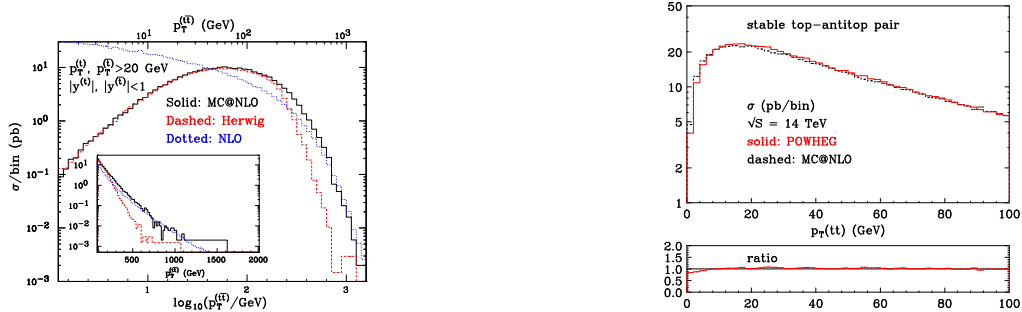}}}
\caption{Transverse momentum of the $t\bar t$ pair at the LHC, according
to HERWIG, MC@NLO and the NLO calculation (left), and according to POWHEG and MC@NLO,
for stable top quarks (right).}  
\label{hwmc}\end{figure}
\begin{figure}[ht!]
\centerline{\resizebox{0.35\textwidth}{!}{\includegraphics{y1_MC_LHC.eps}}%
\hfill%
\resizebox{0.35\textwidth}{!}{\includegraphics{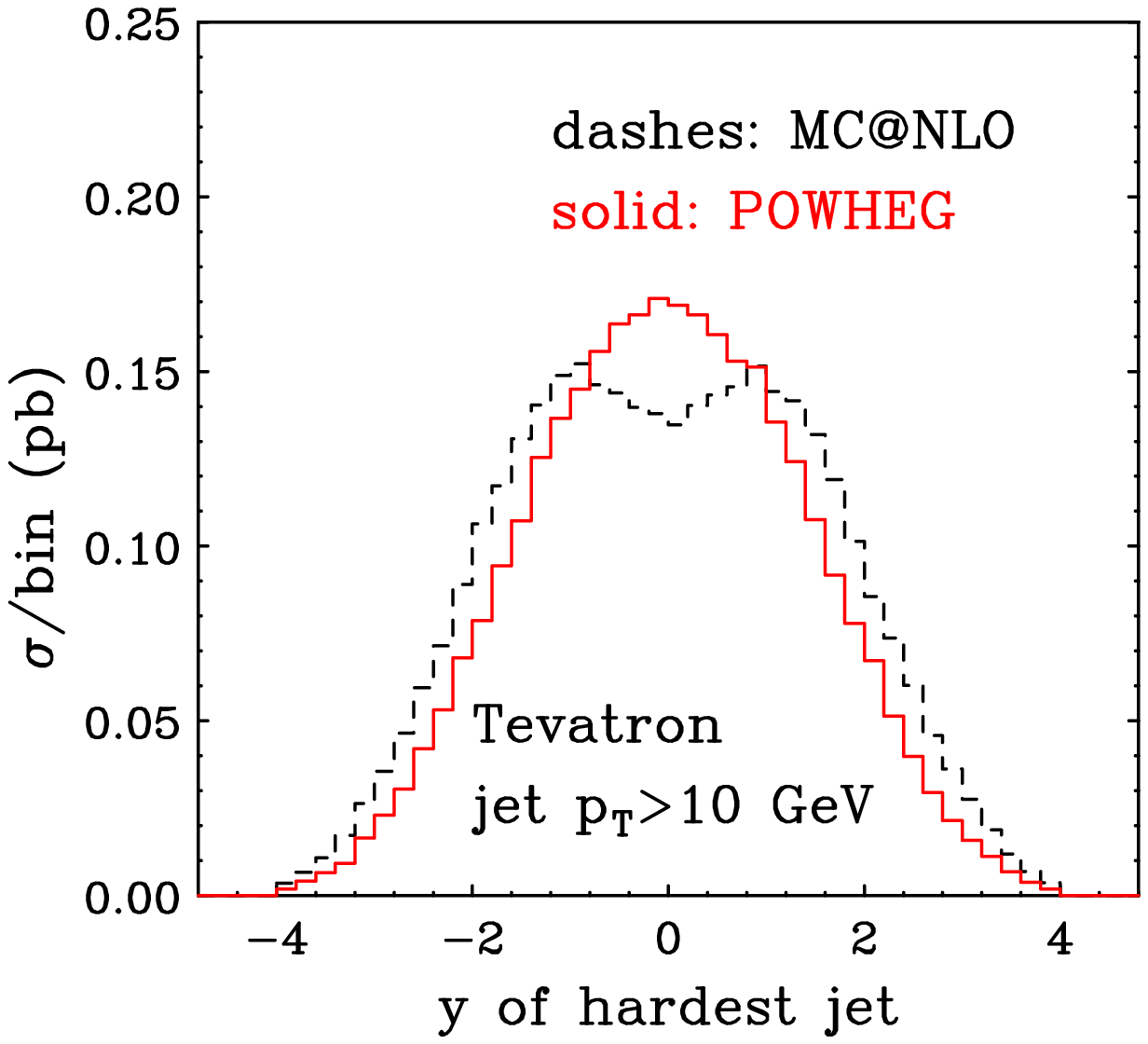}}}
\caption{Rapidity of the hardest jet in $t\bar t(g)$ events at the LHC according
to ALPGEN and MC@NLO (left), and at the Tevatron according to POWHEG and
MC@NLO (right).}\label{y1}
\end{figure}
\par\section{Results for top-quark signals and backgrounds}
We would like to present some results for top-quark signals and backgrounds at the
LHC, using the programs described above.
Following \cite{hwmcnlo}, 
in Fig.~\ref{hwmc} we present the transverse momentum of the $t\bar t$ pair 
at the LHC according to 
HERWIG, MC@NLO and the parton-level NLO calculation, with all
distributions normalized to the same area. 
At small $p_T^{(t\bar t)}$, where one is mostly sensitive to soft/collinear
parton radiation, MC@NLO and HERWIG agree, while the pure NLO calculation is
far above the two codes. At large $p_T^{(t\bar t)}$, it is instead the NLO
computation which is reliable: MC@NLO and the NLO prediction agree,
while HERWIG underestimates the large-$p_T^{(t\bar t)}$ rate.
In Fig.~\ref{hwmc} we also compare MC@NLO and POWHEG
for stable top-quark production: up to a small discrepancy at very low
$p_T^{(t\bar t)}$, the agreement between the two codes is remarkable \cite{mcpo}. 
In Fig.~\ref{y1} we present the rapidity of the hardest jet $y_1$, 
yielded by MC@NLO and ALPGEN: we have clear disagreement around
$y_1=0$, with ALPGEN giving more events than MC@NLO. 
As pointed out in \cite{alpmcnlo},
this is due to the fact that MC@NLO does not generate enough hard scatterings with
$t\bar tg$ in the final state. 
This is confirmed by the fact that POWHEG,
whose first emission is always generated at NLO, 
yields a $y_1$-spectrum similar to ALPGEN (see Fig.~\ref{y1}). 
\begin{figure}[ht!]
\centerline{\resizebox{0.36\textwidth}{!}
{\includegraphics{ptrelj1_tch_tev.eps}}%
\hfill%
\resizebox{0.36\textwidth}{!}{\includegraphics[angle=90]{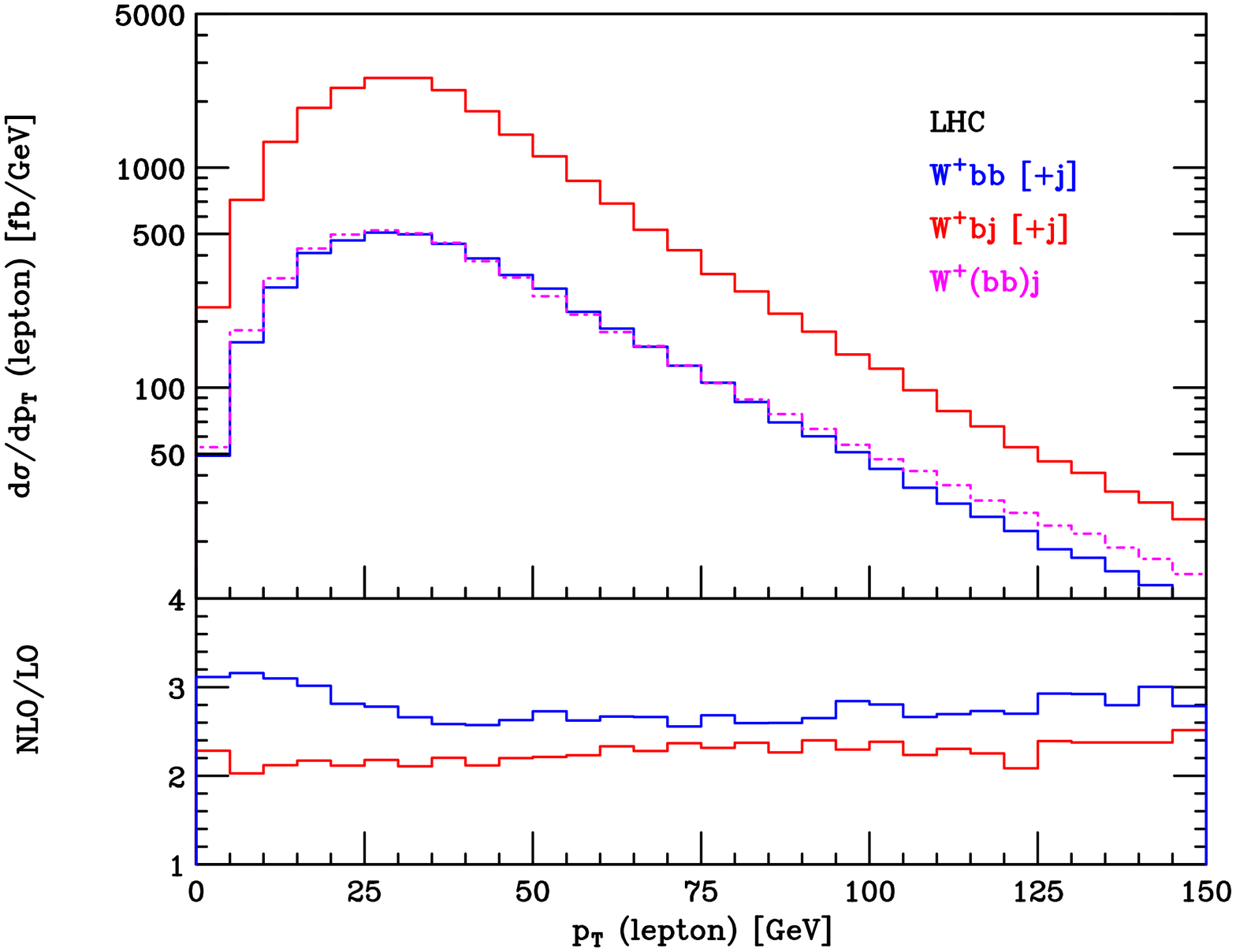}}}
\caption{Left: relative transverse momentum of the hardest-jet
partons with respect to the jet axis for 
single-top production at
the Tevatron. 
Right: lepton $p_T$ for $W+2$ jets with a $b$ tag, with
the $W$ decaying leptonically, at the LHC. The highest cross section is
yielded by the $Wbj[+j]$ subprocesses (solid),
with $Wbb[+j]$ (solid) and $W(bb)j$ (dashes) giving similar contributions.
The NLO/LO ratio is higher for $Wbb[+j]$.} \label{singlet}
\end{figure}
\begin{figure}[ht!]
\centerline{\resizebox{0.36\textwidth}{!}
{\includegraphics{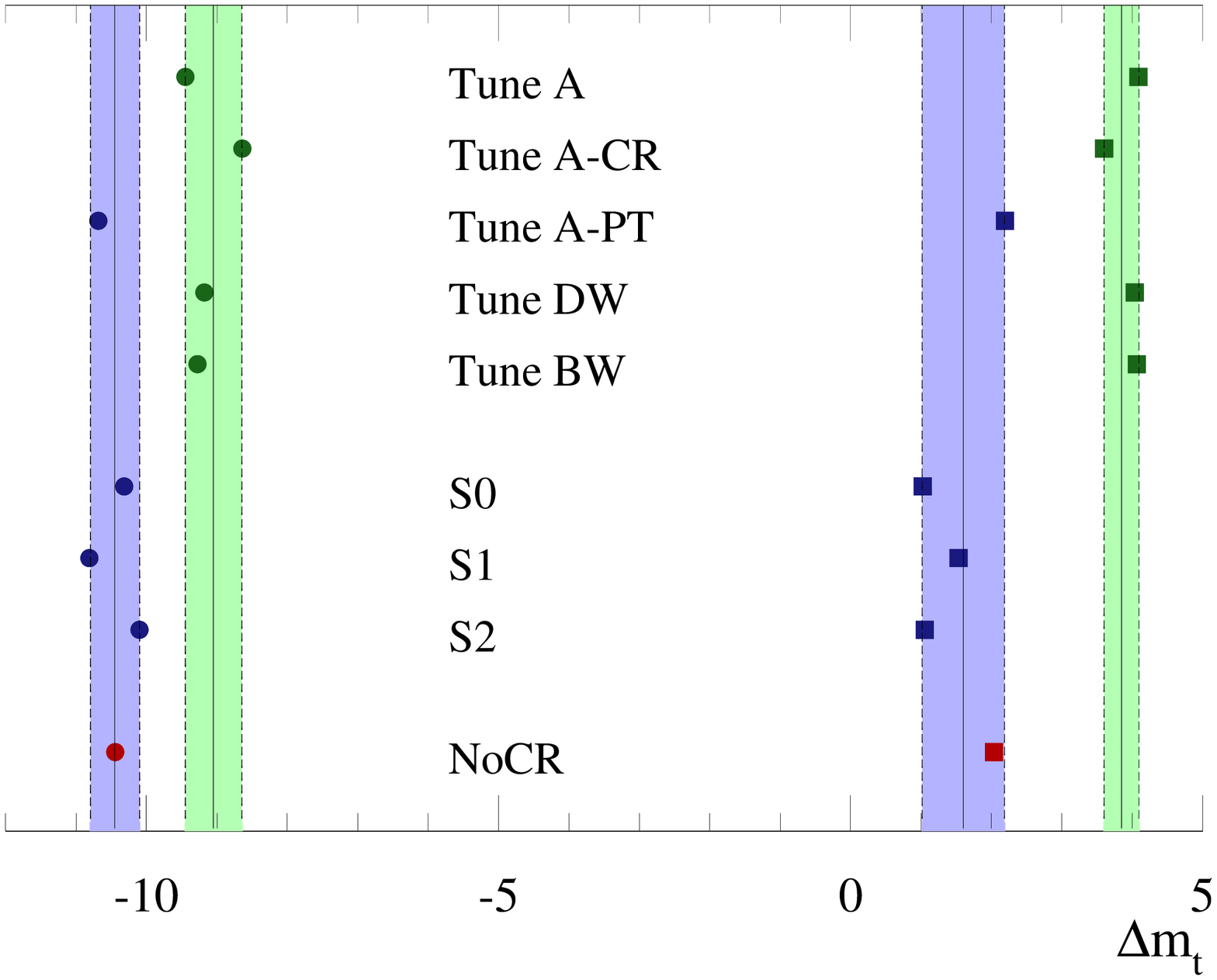}}%
\hfill%
\resizebox{0.36\textwidth}{!}{\includegraphics[angle=90]{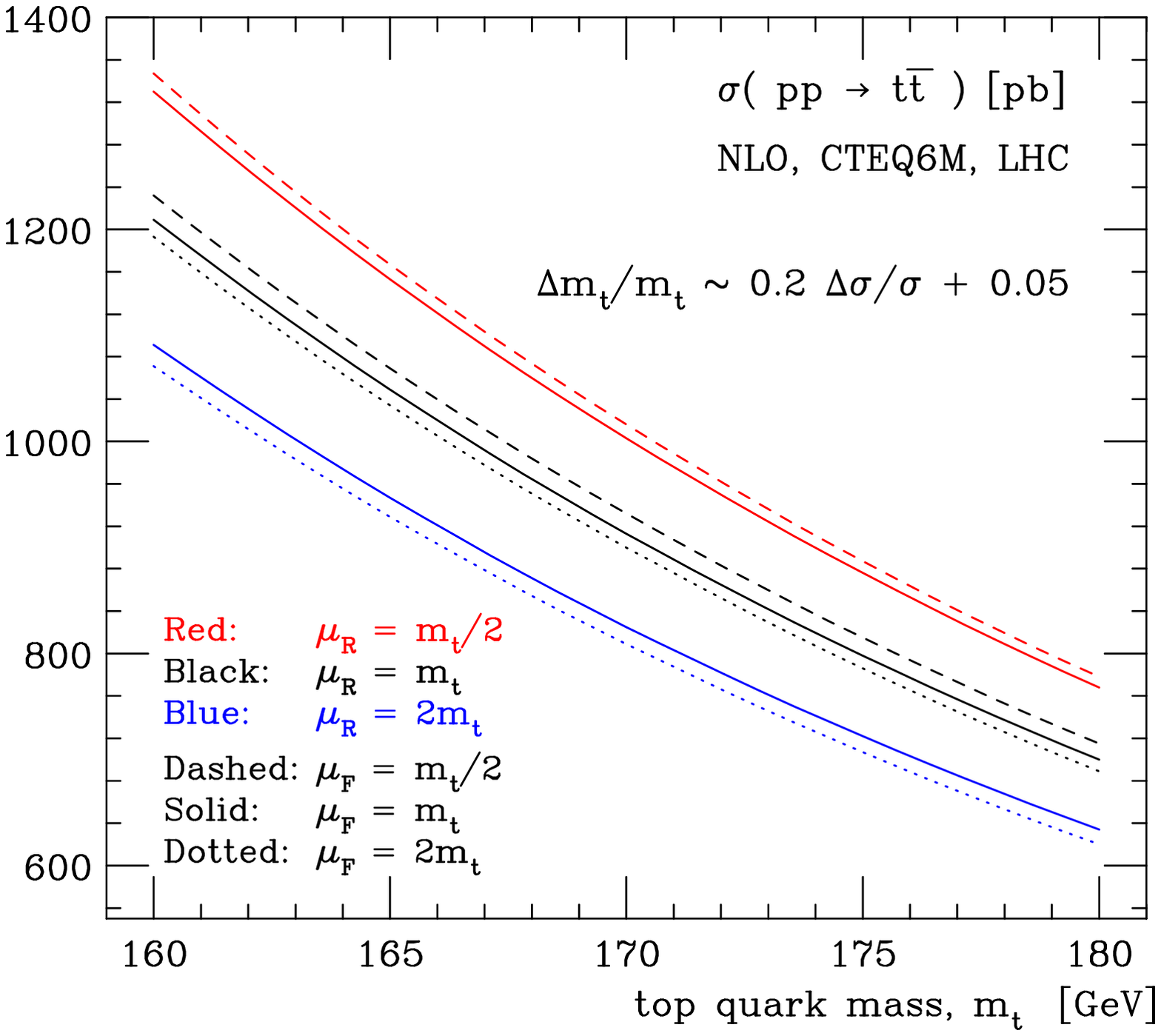}}}
\caption
{Left: uncertainty $\Delta m_t$ due to the underlying event in PYTHIA.
The bands to the
$\Delta m_t>0$ side account for jet-energy-scale corrections.  
In both cases, results on the left- (right-) hand side 
correspond to evolution in transverse momentum (virtuality). 
Right: Dependence of the $t\bar t$ cross section on $m_t$ for few values of
renormalization and factorization scales.
The lines on the top correspond to $\mu_R=m_t/2$, in the middle
to $\mu_R=m_t$, on the bottom to $\mu_R=2m_t$.}\label{und}
\end{figure}
\par For single-top production, a comparison among some of the available
codes was performed in \cite{single}. Fig.~\ref{singlet} shows the total
transverse 
momentum of all partons in the hardest jet, relative to the jet axis, 
using HERWIG, MC@NLO and
the NLO calculation. The agreement between HERWIG and MC@NLO is acceptable
over the full range: in fact, $p^{h}_{T\mathrm{rel}}
$ is a sufficiently inclusive observable,
so that NLO effects mainly result in an overall $K$-factor.
The NLO result is sharply peaked at
$p^{h}_{T\mathrm{rel}}=0$, since at NLO a jet often coincides with a single parton.
At large $p^{h}_{T\mathrm{rel}}$, the NLO result overestimates the prediction of the
two MC codes, where a large-$p^{h}_{T\mathrm{rel}}$ hadron 
likely belongs to another
jet.  Among the backgrounds to single-top production,
$W+2$ jets, at least one with a $b$ quark, were
studied in \cite{camp}, making use of MCFM \cite{mcfm}, a NLO parton-level code.
Fig.~\ref{singlet} shows the $p_T$ distribution of the lepton from $W$ decay,
for the $Wb\bar b+X$,
$Wbj+X$ and $W(b\bar b)j$ processes. At the LHC, $Wbj+X$ is dominant, while the
other two backgrounds are roughly comparable.
\begin{figure}[ht!]
\centerline{\resizebox{0.42\textwidth}{!}
{\includegraphics{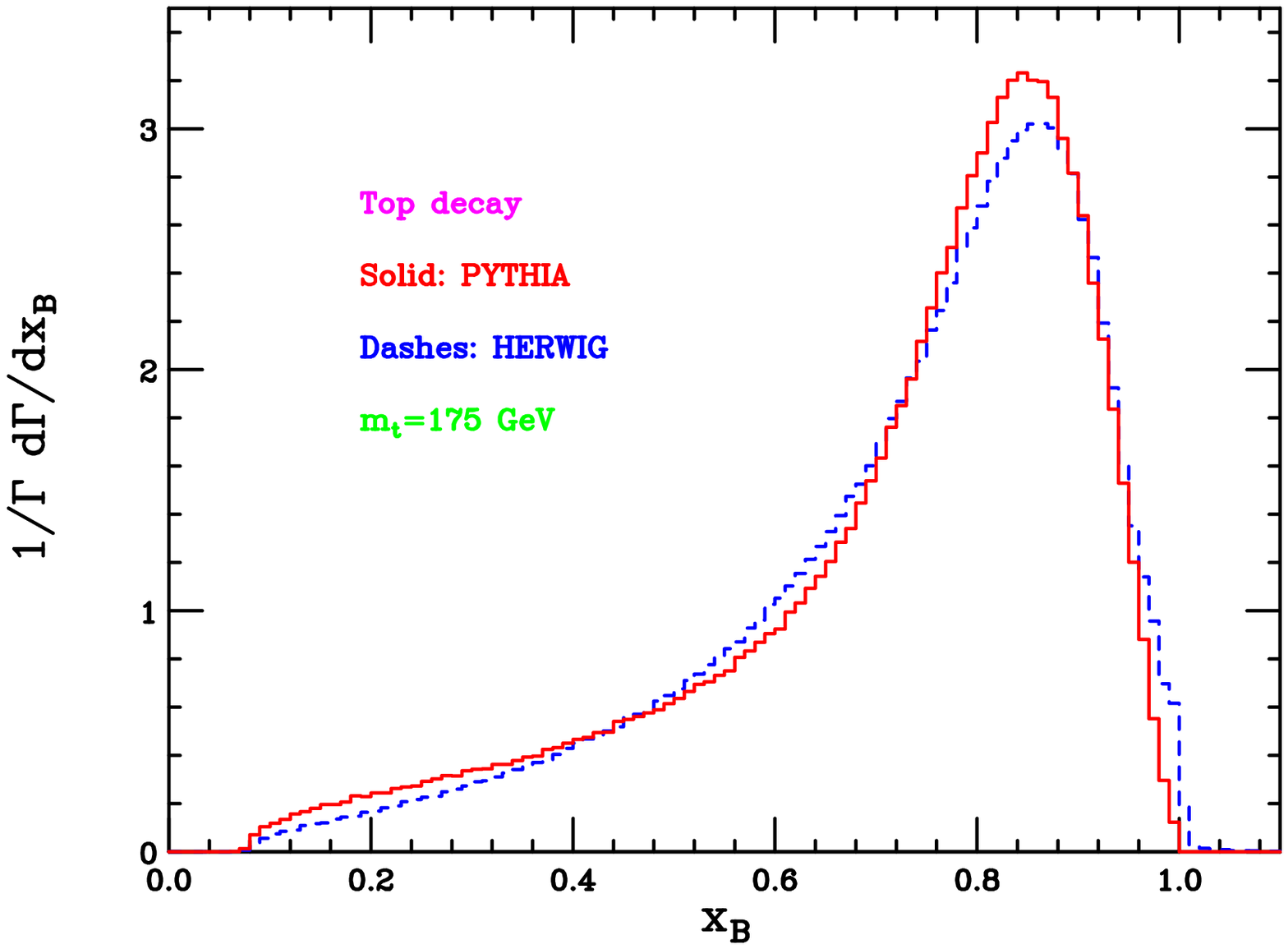}}%
\hfill%
\resizebox{0.42\textwidth}{!}{\includegraphics{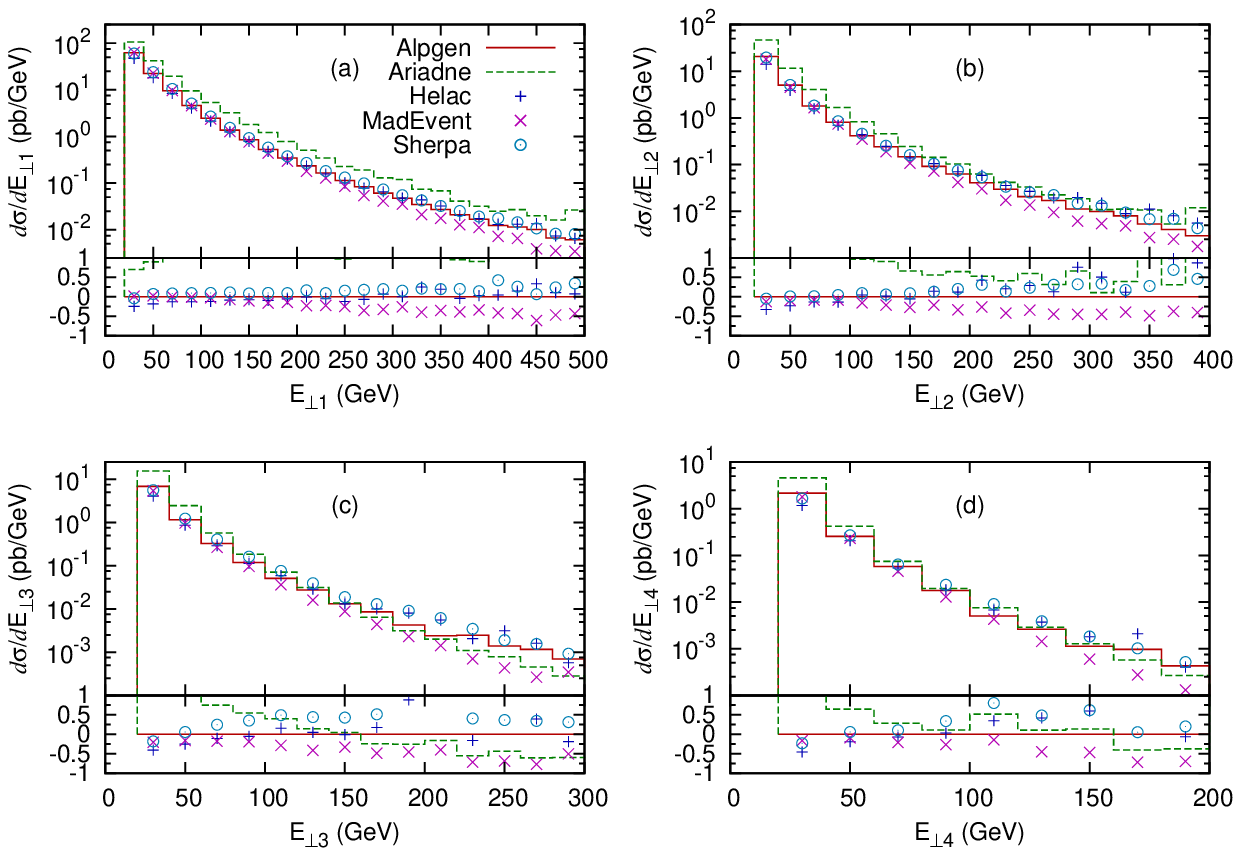}}}
\caption{Left: $B$-hadron spectrum in top decay yielded by 
HERWIG and PYTHIA,
after fitting their hadronization models to LEP and SLD data.
Right: Transverse energy of the first four jets in $W+4$ jet 
events according to a few generators.} 
\label{bfr}
\end{figure}
\begin{table}[b!]
\small
\caption{Cross sections in pb for $(Z/\gamma^*\to e^+e^-)+n$ jets at the LHC according to
different codes. The following cuts are set on jets: $p_{T}>20$~GeV, 
$|\eta|<2.5$,
$\Delta R=\sqrt{(\Delta\phi)^2+(\Delta\eta)^2}>0.4$.}
\begin{center}
\begin{tabular}{| c || c | c | c | c | c | c | c ||}
\hline
$Z+n$~jets & 0 & 1 & 2 & 3 & 4 & 5 & 6  \\
\hline
ALPGEN & 723.4(9) & 188.3(3) & 69.9(3) & 27.2(1) & 10.95(5) & 4.6(1)&
1.85(1)   \\
\hline
SHERPA & 723.9(7) & 189.6(9) & 71.4(4) & 30(2) & & & \\
\hline
CompHEP & 730.9(1) & 190.20(7) & 70.22(7) &  &  & & \\
\hline
MadEvent & 723(1) & 188.6(3) & 69.3(1) & 27.1(2) & 10.6(1) & & \\
\hline
GR@APPA & 744(7) & 182.77(8) & 67.70(3) &  &  & & \\
\hline
\end{tabular}
\end{center}\label{tabjet}
\end{table}  
\par We wish to present a few studies on the theoretical uncertainty
on the top mass measurement. Ref.~\cite{skands} discusses the error 
in the lepton+jets channel, due to different parametrizations of
the underlying event in PYTHIA, with the possible
inclusion of jet-energy-scale corrections. As showed in Fig.~\ref{und}, several
tunings and models are 
investigated and the overall uncertainty is estimated to be 
$\Delta m_t\simeq \pm 0.5$~GeV. 
Scale and parton-density uncertainties are instead
studied in \cite{maltoni}. Fig.~\ref{und} also
presents the NLO cross section $\sigma(pp\to t\bar t)$, as
a function of $m_t$, for various values of renormalization and factorization
scales. The relation $\Delta m_t/m_t\sim 0.2 \ \Delta \sigma/\sigma+0.05$ 
emphasizes that the relative error on $m_t$ cannot be below $5\%$, 
if $m_t$ is determined from a measurement of the cross section.
Another source of uncertainty on $m_t$ is $b$-quark fragmentation in top
decay, which relies on the hadronization 
models implemented in MC generators, fitted
to $e^+e^-$ data. In Ref.~\cite{drol} it was found 
that the default parametrizations
of cluster and string models are uncapable of fitting $B$-hadron data from
LEP and SLD. After fitting few parameters, PYTHIA gives
a good description of the data, whereas HERWIG is only marginally
consistent. Therefore, as shown in Fig.~\ref{bfr}, 
HERWIG and PYTHIA predictions for the
$B$-spectrum in top decay look quite different:
such a result will have an impact on the MC uncertainty on $m_t$.
The analysis \cite{mike} investigates instead the 
systematic error due to initial- (ISR), 
final-state radiation (FSR), underlying event (UE)
and hadronization on the reconstruction of $m_t$ as the invariant 
mass of the $Wb$-jet combination, in the lepton+jets channel.
The $b$-jet is clustered according to the 
$k_T$ (KtJet) and cone (PxCone) algorithms.
Ref.~\cite{mike} found that, relying on
the $k_T$ algorithm, the reconstructed $m_t$ shows 
visible dependence on FSR, ISR and UE, whereas hadronization effects are
very little. When using the cone one,
FSR and hadronization have a large impact, while ISR and UE 
effects are negligible.
Given the different features
of the two algorithms, using both of them is therefore advisable.

As for matrix-element generators, Table~\ref{tabjet} \cite{fulvio} 
quotes the results of a few codes on
$Z+n$ jets, with $n\leq 6$ and $Z\to e^+e^-$, one of the backgrounds for
top production. It is remarkable that, as long
as a process is implemented, the considered codes agree.  More exclusive
studies on $W+$ jets were carried out in \cite{alw},
but even there the used programs
were found to agree. Fig.~\ref{bfr} shows
the transverse energy distribution
of the first four hardest jets
yielded by ALPGEN, ARIADNE, HELAC, MadEvent and SHERPA.

In summary, we reviewed some of the existing event generators for
top signals and background and presented a few results for the LHC.
Given the large numbers of available tools, it is always advisable
using at least two codes for comparison.



\end{document}